\documentclass[aps,prl,twocolumn,superscriptaddress,floatfix]{revtex4}
\usepackage{amsmath}
\usepackage{amsfonts}
\usepackage{amssymb}
\usepackage{graphicx}
\usepackage{cancel}
\usepackage[switch,columnwise]{lineno}
\usepackage{titlesec}
\usepackage{xcolor}
\usepackage{hyperref}

\makeatletter
\renewcommand{\section}{\@startsection{section}{1}{0mm}
  {-\baselineskip}{0.5\baselineskip}{\bf\leftline}}

\renewcommand{\subsection}{\@startsection{section}{1}{0mm}
  {-\baselineskip}{0.5\baselineskip}{\bf\leftline}}
\makeatother

\begin{document}

\title{Non-reciprocity and quantum correlations of light transport in hot atoms via reservoir engineering}

\author{Xingda Lu}%
\affiliation{Department of Physics, State Key Laboratory of Surface Physics and Key Laboratory of Micro and Nano Photonic Structures (Ministry of Education), Fudan University, Shanghai 200433, China}%
\author{Wanxia Cao}%
\affiliation{Department of Physics, State Key Laboratory of Surface Physics and Key Laboratory of Micro and Nano Photonic Structures (Ministry of Education), Fudan University, Shanghai 200433, China}%
\author{Wei Yi}%
\affiliation{CAS Key Laboratory of Quantum Information, University of Science and Technology of China, Hefei 230026, China}
\affiliation{CAS Center For Excellence in Quantum Information and Quantum Physics, Hefei 230026, China}
\author{Heng Shen}%
\email{heng.shen@physics.ox.ac.uk}
\affiliation{State Key Laboratory of Quantum Optics and Quantum Optics Devices, Institute of Opto-Electronics, Shanxi University, Taiyuan 030006, China}
\affiliation{Collaborative Innovation Center of Extreme Optics, Shanxi University, Taiyuan 030006, China}
\affiliation{Clarendon Laboratory, University of Oxford, Parks Road, Oxford, OX1 3PU, UK}%
\author{Yanhong Xiao}%
\email{yxiao@fudan.edu.cn}
\affiliation{State Key Laboratory of Quantum Optics and Quantum Optics Devices, Institute of Laser Spectroscopy, Shanxi University, Taiyuan 030006, China}
\affiliation{Collaborative Innovation Center of Extreme Optics, Shanxi University, Taiyuan 030006, China}
\affiliation{Department of Physics, State Key Laboratory of Surface Physics and Key Laboratory of Micro and Nano Photonic Structures (Ministry of Education), Fudan University, Shanghai 200433, China}%

\begin{abstract}
The breaking of reciprocity is a topic of great interest in fundamental physics and optical information processing applications. We demonstrate non-reciprocal light transport in a quantum system of hot atoms by engineering the dissipative atomic reservoir. Our scheme is based on the phase-sensitive light transport in a multi-channel photon-atom interaction configuration, where the phase of collective atomic excitations is tunable through external driving fields. Remarkably, we observe inter-channel quantum correlations which originate from interactions with the judiciously engineered reservoir. The non-reciprocal transport in a quantum optical atomic system constitutes a new paradigm for atom-based, non-reciprocal optics, and offers opportunities for quantum simulations with coupled optical channels.
\end{abstract}
\maketitle

{\it Introduction--}
Signal transport between two different nodes is a fundamental building block for optical systems. Commonly symmetric between the nodes, such transport can be made directional in non-reciprocal devices such as isolator~\cite{Jalas}, circulator and directional amplifier~\cite{Amp1, Amp2, Amp3}, which undergoes growing interest and demand recently due to their potential utilities in signal processing and quantum networks~\cite{network1,network2}.
Conventionally, non-reciprocity is realized by incorporating magnetized materials which are not limited by the Lorentz reciprocity theorem.
Alternative magnetic-field-free schemes for non-reciprocal light transport include time modulation~\cite{Sounas}, artificial gauge fields~\cite{Painter, DongCH}, reservoir engineering~\cite{Painter, Clerk}, as well as various other approaches~\cite{FanSH,GongSQ,GongSQ2,Kippenberg,TCZhang,YCLiu}. However, despite its importance for quantum information science and applications, the study of non-reciprocity in systems with demonstrated quantum properties has so far been limited to a handful of physical platforms, wherein the quantum correlations rely on the preparation of quantum states of atoms or photons~\cite{Quantum,Yanbo,Singlephoton}.

In this work, we demonstrate non-reciprocal light transport in a quantum system of hot atoms, where the non-reciprocity and quantum correlations derive from reservoir engineering and spin wave interference. Building upon the non-Hermitian platform we developed earlier~\cite{antiPT,Cao}, our setup consists of an array of optical channels immersed in an ensemble of hot atoms, with dissipative, inter-channel couplings mediated by atoms outside the regions illuminated by the lasers. We identify the unilluminated region as a non-Markovian reservoir, whose memory is determined by the lifetime of the ground-state atomic spins. In contrast to previous studies~\cite{GongSQ,GongSQ2,Ming-Xin} of non-reciprocal transport under electromagnetically induced transparency (EIT) where the Doppler-broadening of the EIT linewidth plays a key role, we investigate light transport between two spatially separated optical channels, where the Doppler effect is irrelevant to the observed transport properties.
Furthermore, the phase of the spin excitations as well as parameters of the third, ancillary optical channel are all tunable, thus providing ample tools for reservoir-engineering. A prominent feature of the current configuration is the sensitive dependence of inter-channel light transport on the optical phases, which significantly impact the dissipative coupling between the optical channels and spin excitations in the reservoir. Subsequent manipulation of optical phases, particularly that in the ancillary channel, enables non-reciprocal light transport. We further identify quantum correlations between the two transport channels, which only exist in the presence of the ancillary channel in the reservoir. The observed quantum correlations here derive from the engineered reservoir, fundamentally different from previous works~\cite{antiPT,Cao} where quantum correlations were established via inter-channel couplings mediated by an unstructured environment.

\begin{figure}
\centering
\includegraphics[width=0.43\textwidth]{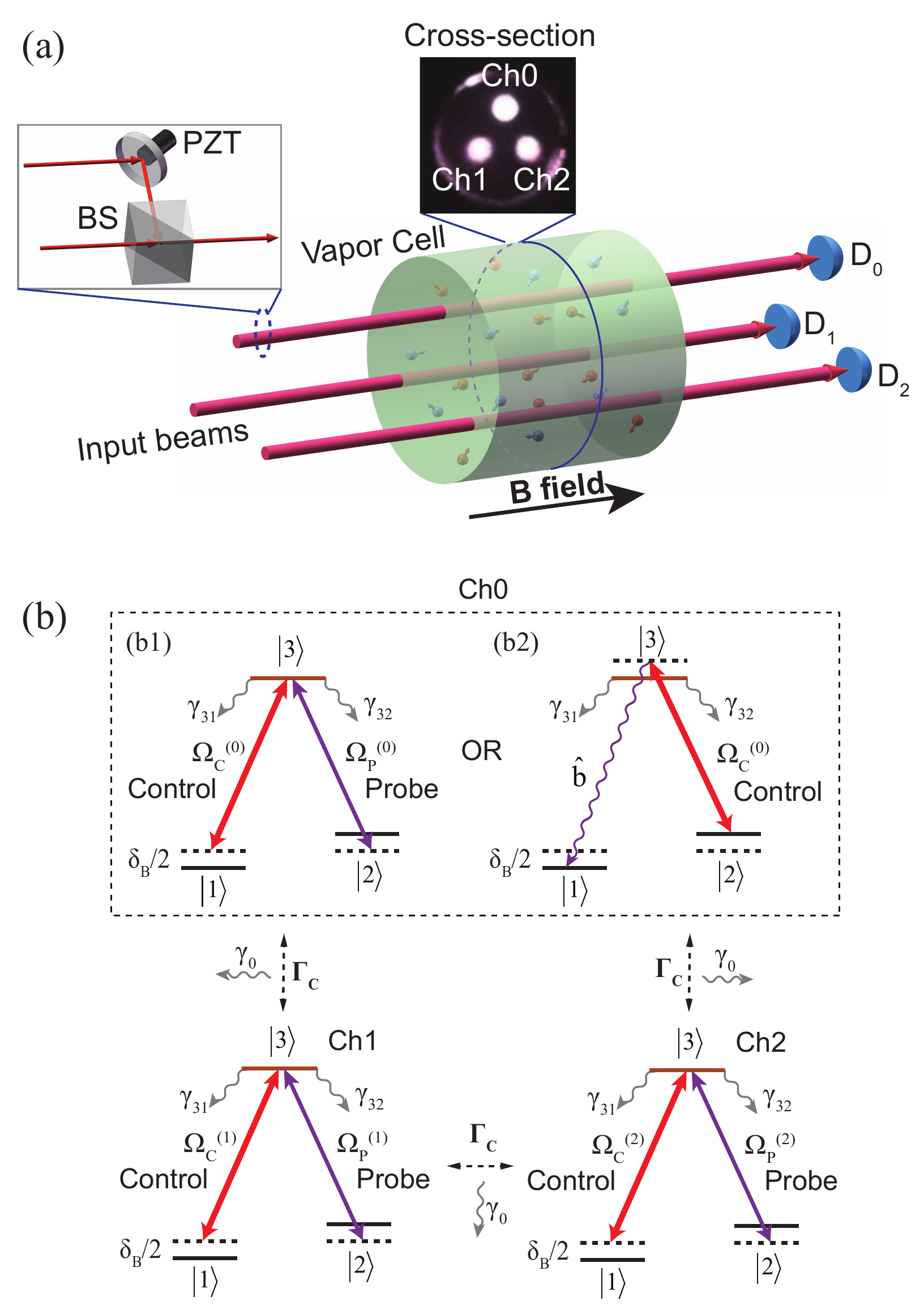}
\caption{\label{Fig:setup} Schematics for a dissipatively coupled three-channel system. (a) Experiment schematics. Three spatially separated optical channels (Ch0, Ch1 and Ch2 with diameter 6 mm) propagate in a warm paraffin-coated $^{87}\text{Rb}$ vapor cell under EIT interaction. The inter-channel couplings are mediated by the mixing of atomic spin of the ground states through atomic motion. A solenoid gives precise control over the longitudinal magnetic field. Output beams from the cell are re-collimated and detected by the photon detectors or polarization homodyne detection setup. The noise power of the amplified subtracted photocurrents is recorded with a spectrum analyzer. BS, beam splitter; $D_0$, $D_1$, and $D_2$ photodetectors. Cross section: photo of the optical beams taken from this experiment. (b) The $\Lambda$ three-level scheme in three channels. The ground states are Zeeman sublevels of $\left |F=2 \right \rangle$, and the excited state is $\left |F=1 \right \rangle$ of the $^{87}\text{Rb}$ \textit{D1} line. $\Omega_{c}^{(i)}$, $\Omega_{p}^{(i)}$, $i=0,1,2$, are Rabi frequencies of the control and probe beams respectively. The Zeeman splitting is induced by a common longitudinal magnetic field $\delta_{B}$, serving as either the Larmor frequency in the noise spectra measurement or the two-photon detuning in the EIT measurement (denoted as $\delta_B$ in Fig.~\ref{Fig:non-reciprocity}). In the measurements of quantum fluctuation, all three weak probes are removed as shown in (b2)
(Ch1 and Ch2 are not shown). $\hat{b}$, annihilation operator of the coherent vacuum.}
\end{figure}

{\it Experimental setup--}
The system under consideration is illustrated schematically in Fig.~\ref{Fig:setup}. A triangular array of spatially-separated optical beams (channels), labeled Ch1, Ch2 and Ch0, propagate in a warm atomic vapor cell. Each channel undergoes a $\Lambda$-type EIT interaction with the same configuration (Fig.~\ref{Fig:setup}(b1)), where a relatively strong control and a weak probe fields conspire to create a collective spin wave $\rho_{12}$ (ground-state coherence) in the atomic ensemble. We apply co-propagating probe and control beams, such that the EIT linewidth is not Doppler-broadened.
While EIT creates a linear mapping between the probe light and the spin excitation~\cite{Fleischhauer,YZH}, spin waves in different channels couple to one another with a rate $\Gamma_c$ via random atomic motion and wall bouncing~\cite{antiPT,Ofer_Eugene}. Although the random atomic motion renders the coupling dissipative, optical phases imprinted upon the atomic spin states are preserved throughout the dynamics, and constitute a convenient control over the light transport.
To study light transport, we take Ch1 and Ch2 as the two optical nodes, and regard Ch0 as part of the reservoir, together with the surrounding atomic ensemble. Since Ch1 and Ch2 couple to the same reservoir, we can manipulate the coherence in the reservoir through Ch0, and thus control the dissipative inter-channel couplings.

As outlined in Fig.~\ref{Fig:setup}(b), Ch1, Ch2 and Ch0 couple to each other with the same underlying mechanism. For any two channels, for instance, Ch1 and Ch2, the control and probe beams in Ch1 write a collective internal-state coherence into the atomic ensemble, which is then read out by the control beam in Ch2, resulting in a non-Hermitian beam-splitter (BS) type interaction~\cite{SlowBS,antiPT} dictated by $\hat{H}\propto\hat{a}_1^{\dagger}\hat{a}_{2}e^{i\psi_{1}}-\hat{a}_1\hat{a}_{2}^{\dagger}e^{-i\psi_{1}}$. Here $\hat{a}_1$ and $\hat{a}_2$ are the annihilation operator of probe beams in Ch1 and Ch2 respectively, and $\psi_{1}$ is the relative phase between control beams in Ch1 and Ch2~\cite{supp}.

For the experiment, we implement the triangular array in a cylindrical vapor cell (with a diameter of 2.5 cm and length of 7.5 cm) which contains isotopically enriched $^{87}$Rb vapor at an operational temperature of 60$^{\circ}$C. The cell is mounted inside a four-layer magnetic shielding, where a set of coils provide precise control over the internal magnetic field. A diode laser is tuned to the D1 line of $^{87}$Rb. The output of the laser is sent through a polarization-maintaining optical fiber, before it is divided into three channels whose polarizations are separately controlled by a combination of half-wave and quarter-wave plates. In each channel, a control field and a weak (or quantum) probe with overlapping spatial profiles induce an EIT process. The spin dynamics of moving atoms can be described by a set of coupled differential equations which take into account the spin transport between different regions as well as the Langevin noise terms~\cite{supp}. The effective coherence exchange between optical channels is mediated by atoms outside the illuminated interaction regions, whose spin states decay slowly (with a lifetime $~30$ ms) due to the protective wall coating~\cite{Balabas,Borregaard,Firstenberg}. The optical coherence transfer between different channels is negligible as it decays here within 20 ns.

\begin{figure}
\centering
\includegraphics[width=0.5\textwidth]{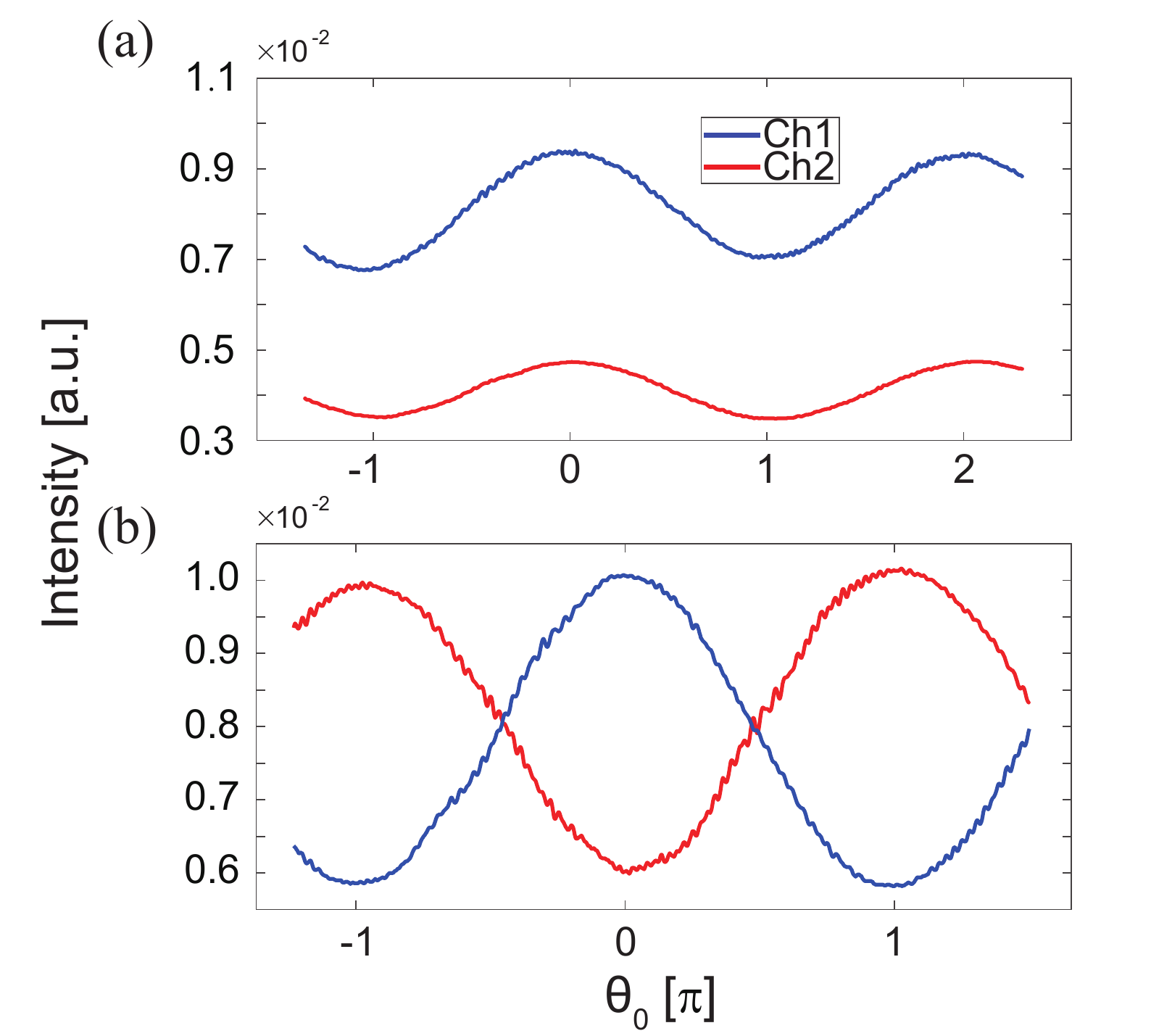}
\caption{\label{Fig:Geometric phase} Phase sensitive light transport for the beam splitter composed of Ch1 and Ch2.
Output probe intensities of Ch1 and Ch2 as functions of varying local phase $\theta_0$ in Ch0, when (a) the weak probe input in Ch2 is off, and (b) all three input probes are on. The input laser powers in all channels are 500 $\mu$W for the control and 50 $\mu$W for the probe respectively, corresponding to Rabi frequencies $\Omega_{c}\sim{1}\times{10^{7}}$ Hz and $\Omega_{p}\sim{4}\times{10^{6}}$ Hz. The local phase of Ch1 is set to be: $\theta_{1}=\psi_{c}^{(1)}-\psi_{p}^{(1)}=0$ and for Ch2: $\theta_{2}=\pi$ (if the probe is on). The cell temperature is 60$^{\circ}$C.}
\end{figure}

{\it Phase-sensitive transport--}
We first investigate the phase dependence of light transport, which is the basis for breaking the reciprocity. We denote the phase of the local spin wave in the $i$th channel as $\theta_i$ ($i=0,1,2$), with  $\theta_{i}=\psi_{c}^{(i)}-\psi_{p}^{(i)}$ where $\psi^{(i)}_c$ and $\psi^{(i)}_p$ are the phases of the control and probe fields in the corresponding channel, respectively. Since inter-channel couplings are mediated by collective spin waves, we have $\psi_{p}^{i\to{j}}=\psi_{c}^{(j)}-\theta_i$, where $\psi_{p}^{i\to{j}}$ denotes the phase of the photons transferred from channel $i$ to channel $j$. The transferred photons then interfere with local probe field, with the resulting interference pattern sensitively dependent on the phase parameters of all channels. Here we focus on the light transport between Ch1 and Ch2, which, as we show below, features non-trivial dependence on $\theta_0$, a tunable parameter of the reservoir.
More specifically, we continuously vary $\theta_{0}$ by slowly sweeping the optical path length of the control beam in Ch0 using a piezoelectric transducer (PZT) (see Fig.~\ref{Fig:setup}(a)), and record the intensities of the weak probe fields in Ch1 and Ch2, respectively.

We start with the case where the probe field in Ch2 is switched off, while the probe field in Ch1 is present with $\theta_1=0$.
The measured probe intensity $I_2$ in Ch2 thus solely derives from the transferred light from Ch1 and Ch0, with $I_{2}\propto|e^{i\psi_{p}^{1\to{2}}}+e^{i\psi_{p}^{0\to{2}}}|^{2}\propto{1+\cos\theta_{0}}$. Likewise, the measured probe intensity in Ch1 is $I_1\propto|1+e^{i\psi_{p}^{0\to 1}}|^2\propto{1+\cos\theta_{0}}$. Crucially, $I_1$ should have the same $\theta_0$ dependence as that of $I_2$, which is confirmed by our experimental measurement in Fig.~\ref{Fig:Geometric phase}(a).

For the second case, we switch on the probe fields in both Ch1 and Ch2, with $\theta_1=0$ and $\theta_2=\pi$, respectively. As shown in Fig.~\ref{Fig:Geometric phase}(b), the phase dependence of detected light intensities in Ch1 and Ch2 deviates drastically. For $\theta_0=\pi$, the detected probe-field intensity in Ch1 is at a minimum, in contrast to the maximum output from Ch2. Following the analysis in the previous case, the output probe-beam intensities are given by $I_1\propto |1+\beta e^{-i\theta_{0}}+\beta e^{-i\pi}|^2$ and $I_2\propto |1+\beta e^{i(\pi-\theta_{0})}+\beta e^{i\pi}|^2$, respectively, where $\beta$ is the beam splitter ratio that can be determined from experimental measurement. As such, the light transport between Ch1 and Ch2 critically depends on the phase parameter $\theta_0$ of the reservoir, owing to the reservoir-mediated interference. The visibility of the oscillations in Fig.~\ref{Fig:Geometric phase} is mainly limited by the relatively low EIT contrast and the small inter-channel coupling rate.

\begin{figure}
\centering
\includegraphics[width=0.5\textwidth]{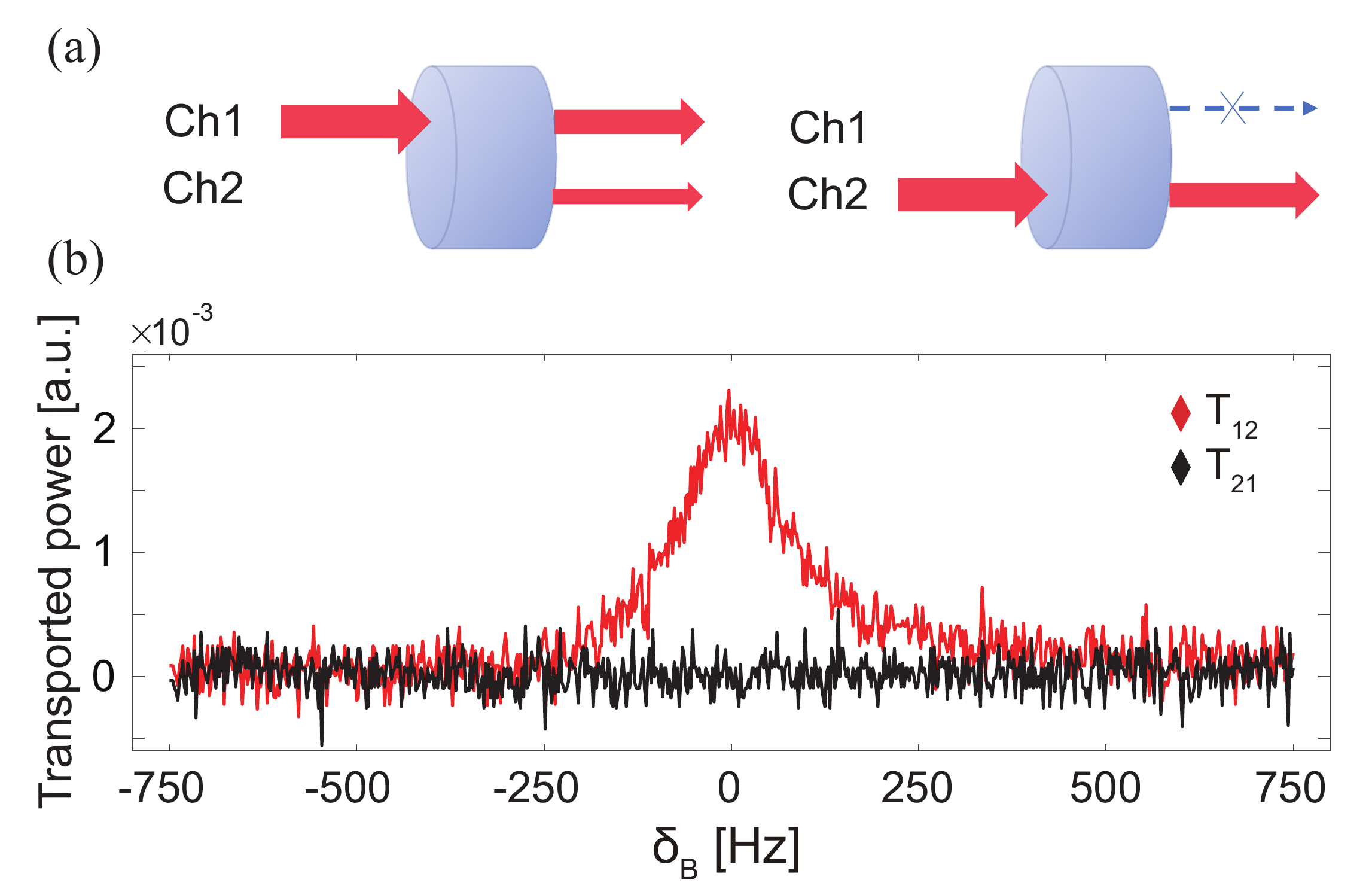}
\caption{\label{Fig:non-reciprocity} Spin wave interference induced non-reciprocity in optical transport. (a) Schematics of the non-reciprocal transport. (b) Transport spectrum. The two-photon detuning $\delta_{B}$ is proportional to the applied common magnetic field. Red curve $T_{12}$ is the transported power from Ch1 to Ch2 when injecting the weak probe in Ch1. Black curve $T_{21}$ is the transported power from Ch2 to Ch1 when injecting the weak probe in Ch2. The local phase of all three channels is set to be: $\theta_{0}=0$, $\theta_{1}=0$ and $\theta_{2}=\pi$. The input power of the probe in each channel is 50 $\mu$W. The input power of the control in each channel is 500 $\mu$W. The cell temperature is 60$^{\circ}$C.}
\end{figure}

{\it Non-reciprocal light transport--}
The reservoir-mediated interference demonstrated above allows the design of non-reciprocal light transport between Ch1 and Ch2 under appropriate reservoir parameters. For instance, we choose the parameters: $\theta_{0}=0$, $\theta_{1}=0$, and $\theta_{2}=\pi$; and measure the EIT spectra by sweeping the applied magnetic field to vary the two-photon detuning as illustrated in Fig.~\ref{Fig:setup}(a). The inter-channel light transport is characterized by switching on the probe field of the input channel (such as Ch1/Ch2), and measuring the light intensity in the output channel (such as Ch2/Ch1) whose probe field is initially switched off. When Ch1 is set as the input channel (left panel of Fig.~\ref{Fig:non-reciprocity}(a)), the EIT spectra in Ch2 has a transmission window near the two-photon resonance $\delta_{B}=0$, where $T_{12}\propto{|e^{i\psi_p^{1\to 2}}+e^{i\psi_p^{0\to 2}}|^2}\propto1+\cos(\theta_{0}-\theta_{1})=2$, peaking under a constructive interference. By contrast, when Ch2 is the input channel (right panel of Fig.~\ref{Fig:non-reciprocity}(a)), the EIT spectra in Ch1 becomes $T_{21}\propto{|e^{i\psi_p^{2\to 1}}+e^{i\psi_p^{0\to 1}}|^2}\propto{1+\cos(\theta_{0}-\theta_{2})=0}$, vanishing due to destructive interference.
The analysis here are consistent with our experimental observation in Fig.~\ref{Fig:non-reciprocity}(b), where the directional light transport is clearly identified. The origin of the observed non-reciprocity is therefore the interference of spin waves along the path of the atomic motion, analogous to the time-modulation scheme~\cite{Jalas}.
The light transport can be easily tuned to be reciprocal, for instance, by setting $\theta_0=\pi/4$. For all our experiments here, the observed isolation is $\sim$19 dB, and the un-transported power is absorbed by the atoms. In the Supplementary Material, we show that the non-reciprocity persists under a bi-directional input, where probes in both channels are switched on.

{\it Quantum correlation--}
Whereas the non-reciprocal light transport demonstrated above relies only on the interference of spin waves, a surprising finding is that quantum correlations in the polarization degree of freedom of light can be established between the transport channels Ch1 and Ch2, as a consequence of interactions with the engineered reservoir.

\begin{figure}
\centering
\includegraphics[width=0.5\textwidth]{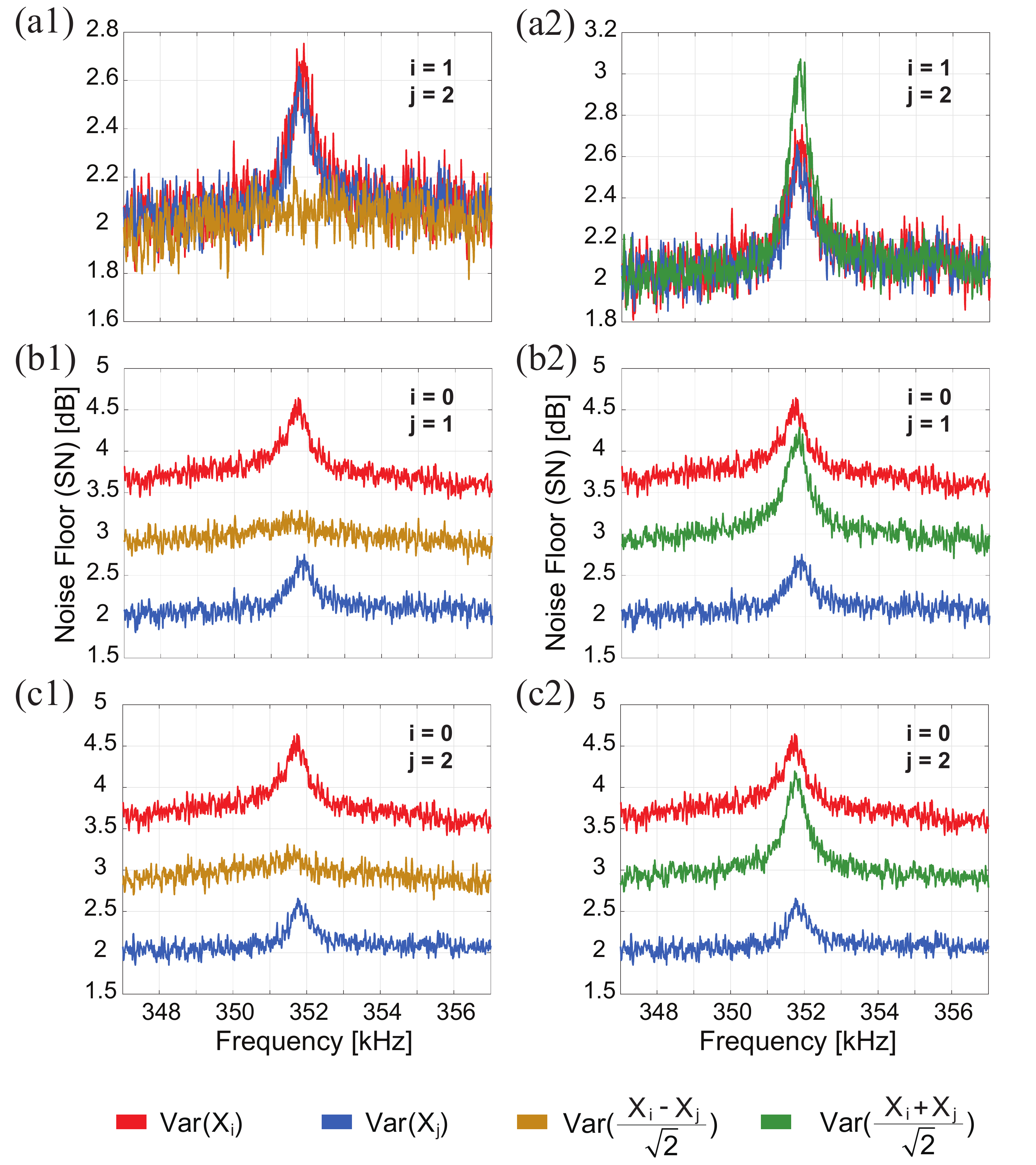}
\caption{\label{Fig:quantum correlation} Quantum correlations exist between any two channels. The experiment noise spectra of Var($\hat{X}_i$), Var($\frac{\hat{X}_i\mp\hat{X}_j}{\sqrt{2}}$) ($i,j=0$, 1, 2 and $i\ne{j}$) are shown. They are identical to Var($\hat{P}_i$), Var($\frac{\hat{P}_i\pm\hat{P}_j}{\sqrt{2}}$) respectively (not shown). The shot noise level is set at 0 dB, observed when only one channel is on. The bias magnetic field is applied to shift the homodyne measurement from DC to the Larmor frequency($\sim$352kHz). The input control power is 280 $\mu$W in both Ch1 and Ch2, and 500 $\mu$W in Ch0. The cell temperature is 60$^{\circ}$C.}
\end{figure}

To demonstrate the presence of quantum correlation, we reverse the polarizations of the control and probe in Ch0 as shown in Fig.~\ref{Fig:setup} (b2). While the phase sensitive, non-reciprocal light transport persists under this new configuration (as we have confirmed experimentally~\cite{supp}), counter-intuitive quantum correlations now emerge.
To facilitate quantum measurements under the constraints of the relatively low optical depth and technical noise in the lasers, we switch off the input probe fields in all three channels, replacing them with coherent vacua.
We define the canonical position and momentum operators of the $i$th channel through Stokes operators: $\hat{X}_i=\hat{S}_{x}^{i}/\sqrt{\left | S_{z}^{i} \right |}$ and $\hat{P}_i=\hat{S}_{y}^{i}/\sqrt{\left | S_{z}^{i} \right |}$. The noise spectra $\text{Var}(\hat{X}_{i})$ and joint variance $\text{Var}(\frac{\hat{X}_i\pm \hat{X}_j}{\sqrt{2}})$ ($i$, $j$=0, 1, 2 and $i\neq j$) of the steady state are then measured via joint polarization homodyne detections~\cite{Cao}. For the detection, a bias magnetic field is applied along the propagation direction of light to shift the homodyne measurement from DC to the Larmor frequency ($\sim$352 kHz), bypassing low-frequency technical noise. As illustrated in Fig.~\ref{Fig:quantum correlation}, quantum correlations manifest as $\text{Var}(\frac{\hat{X}_i-\hat{X}_j}{\sqrt{2}})+\text{Var}(\frac{\hat{P}_i+\hat{P}_j}{\sqrt{2}})<\frac{\text{Var}(\hat{X}_{i})+\text{Var}(\hat{X}_{j})}{2}+\frac{\text{Var}(\hat{P}_{i})+\text{Var}(\hat{P}_{j})}{2}$. To quantify the measured quantum correlation, we calculate the Gaussian discord $\mathfrak{D}_{ij}$~\cite{Adesso,Paris} at the Larmor frequency in the noise spectra between different channels. In the presence of a dissipative environment, Gaussian discord captures Gaussian quantum correlations, and is more robust than quantum entanglement in revealing quantum correlations. Following its definition, we have $\mathfrak{D}_{01}=2.9\times 10^{-3}$, $\mathfrak{D}_{02}=2.5\times 10^{-3}$, and $\mathfrak{D}_{12}=2.5\times 10^{-3}$. The positiveness of the evaluated discords unambiguously indicates the quantum nature of correlation between any two channels. The measured Gaussian discord is relatively low due to the small inter-channel coupling rate in our experiment, as significant information is lost to reservoir. For future studies, lasers with larger beam size and non-Gaussian profiles could be employed to enhance the quantum correlation between the optical channels.

It is worth emphasizing that, the quantum correlation between Ch1 and Ch2 is counter-intuitive and derive purely from the engineered reservoir by Ch0. In the absence of Ch0, i.e., with an unstructured reservoir as in ~\cite{Cao}, the interaction between Ch1 and Ch2 is of the beam-splitter type, and therefore the probe output of Ch1 and Ch2 are simply photon-shot noise, with a vanishing discord $\mathfrak{D}_{12}$. However, in the presence of Ch0, both $\mathfrak{D}_{01}$ and $\mathfrak{D}_{02}$ become finite, due to the two-mode-squeezing (TMS) type interaction between Ch0 and Ch1, as well as between Ch0 and Ch2, due to their opposite polarization configurations. Apparently, the quantum correlation between Ch1 and Ch2 originate from the interplay of the beam-splitter type interaction between Ch1 and Ch2, and their TMS-type interactions with the reservoir containing Ch0.

{\it Conclusion--}
We have introduced a platform with non-reciprocal and quantum transport of light, based on hot-atom vapor cell in a spin-coherence-protected environment. Both the observed non-reciprocal transport and quantum correlations between the optical channels derive from the interference mediated by an engineered reservoir, and are tunable by adjusting parameters of all the optical channels including the one embedded in the reservoir. Our work provides a prototype configuration for an atom-based, non-reciprocal optical element. Based on the geometry of our setup and benefiting from the high degree of control over the atom-light interactions, the configuration demonstrated here may further offer opportunities for quantum simulation in open systems~\cite{Dalibard2,Bloch2} using multiple coupled optical channels.

We thank Mikhail Balabas and Precision Glassblowing (Colorado, USA) for assistance in the vapor cell fabrication. The authors are grateful to Peter Zoller for stimulating discussions. This work is supported by National Key Research Program of China under Grant Nos. 2017YFA0304204, 2016YFA0302000 and 2020YFA0309400, Natural Science Foundation of China (NSFC) under Grant Nos. 61675047 and 12027806. H. Shen acknowledges the financial support from the Royal Society Newton International Fellowship Alumni follow-on funding (AL201024) of United Kingdom. W. Yi acknowledges support from NSFC under Grant No. 11974331, and the National Key Research and Development Program of China under Grant Nos. 2016YFA0301700 and 2017YFA0304100.

\begin{figure}
\centering
\includegraphics[width=1\textwidth]{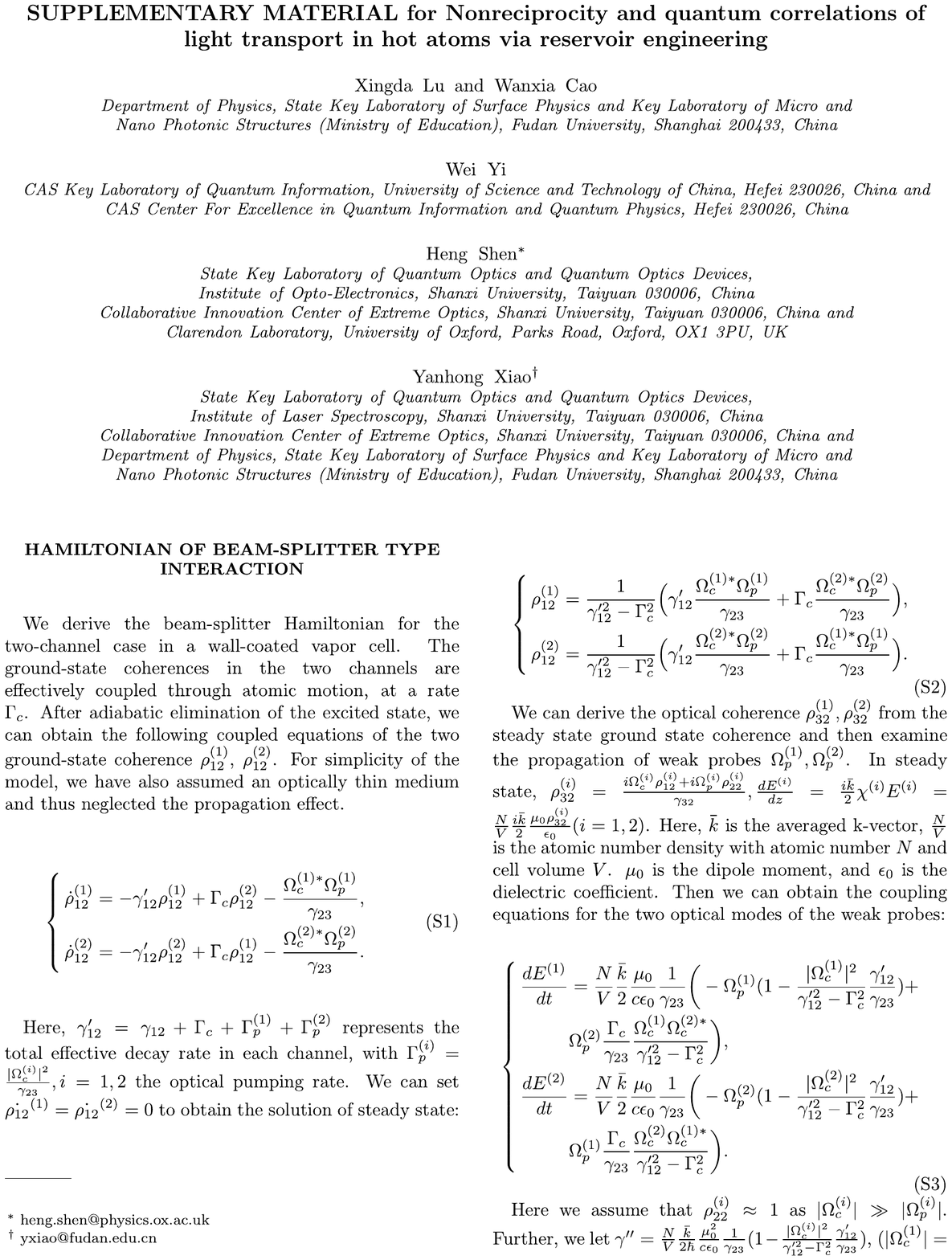}
\end{figure}

\begin{figure}
\centering
\includegraphics[width=1\textwidth]{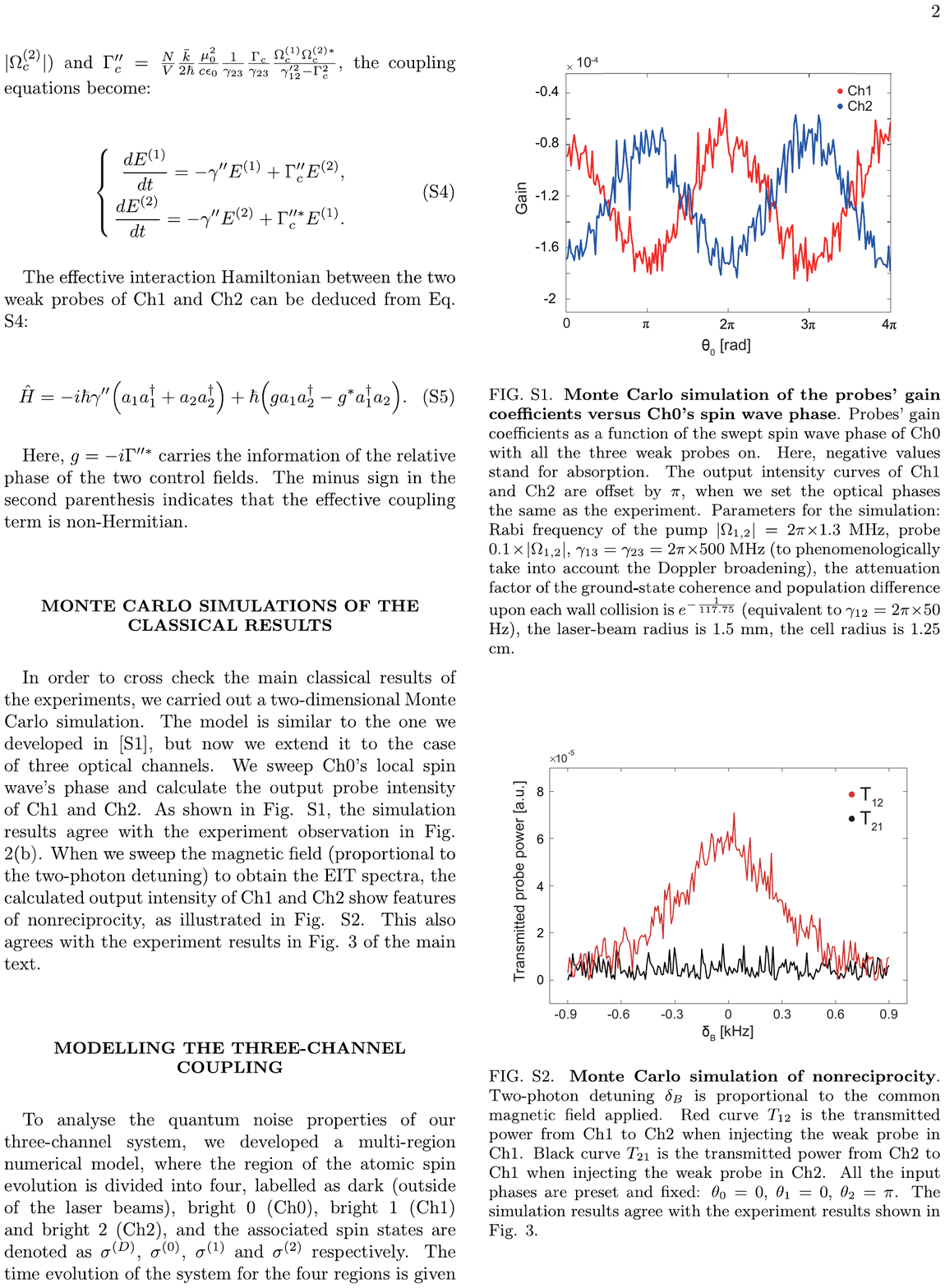}
\end{figure}

\begin{figure}
\centering
\includegraphics[width=1\textwidth]{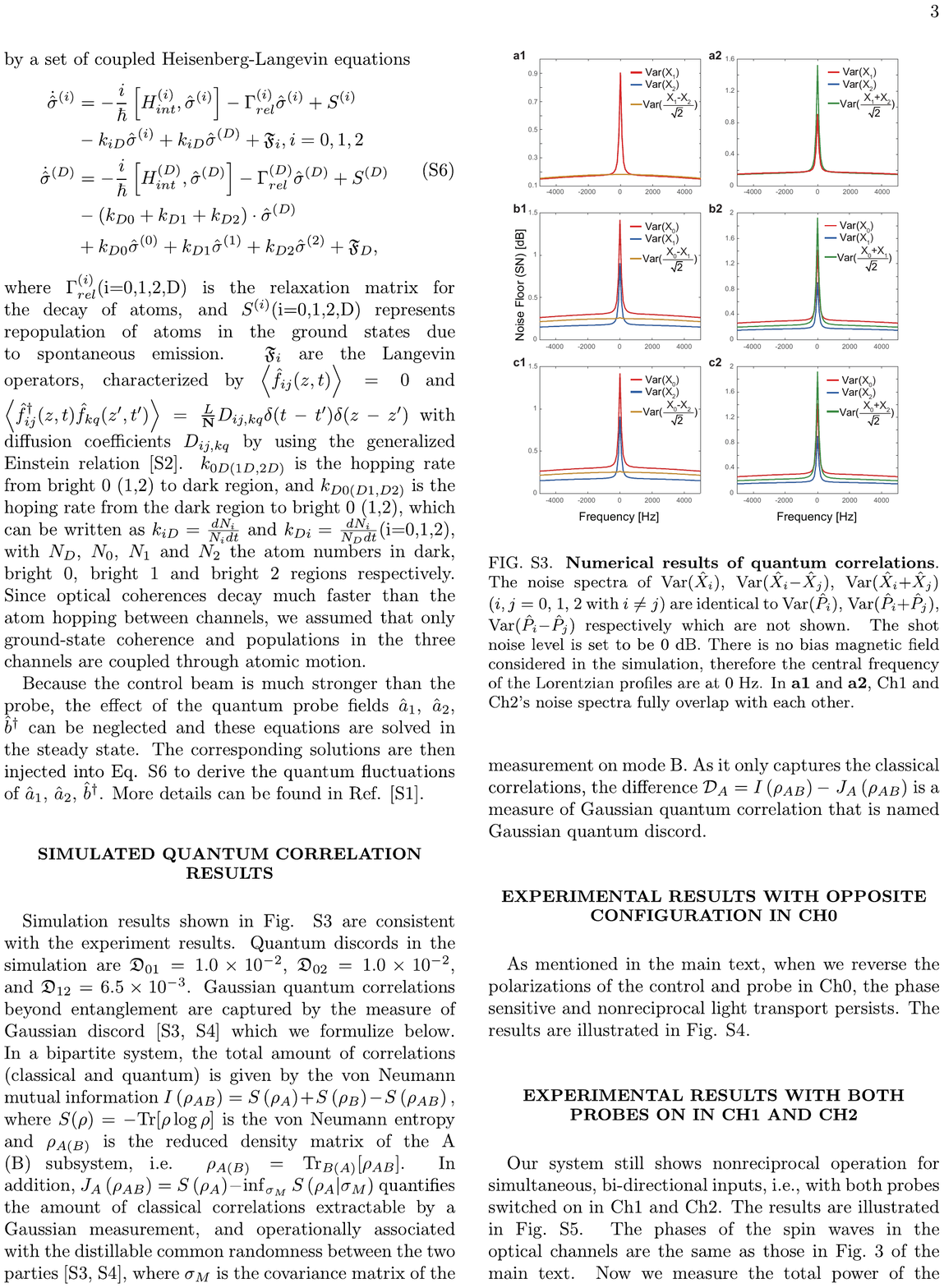}
\end{figure}

\begin{figure}
\centering
\includegraphics[width=1\textwidth]{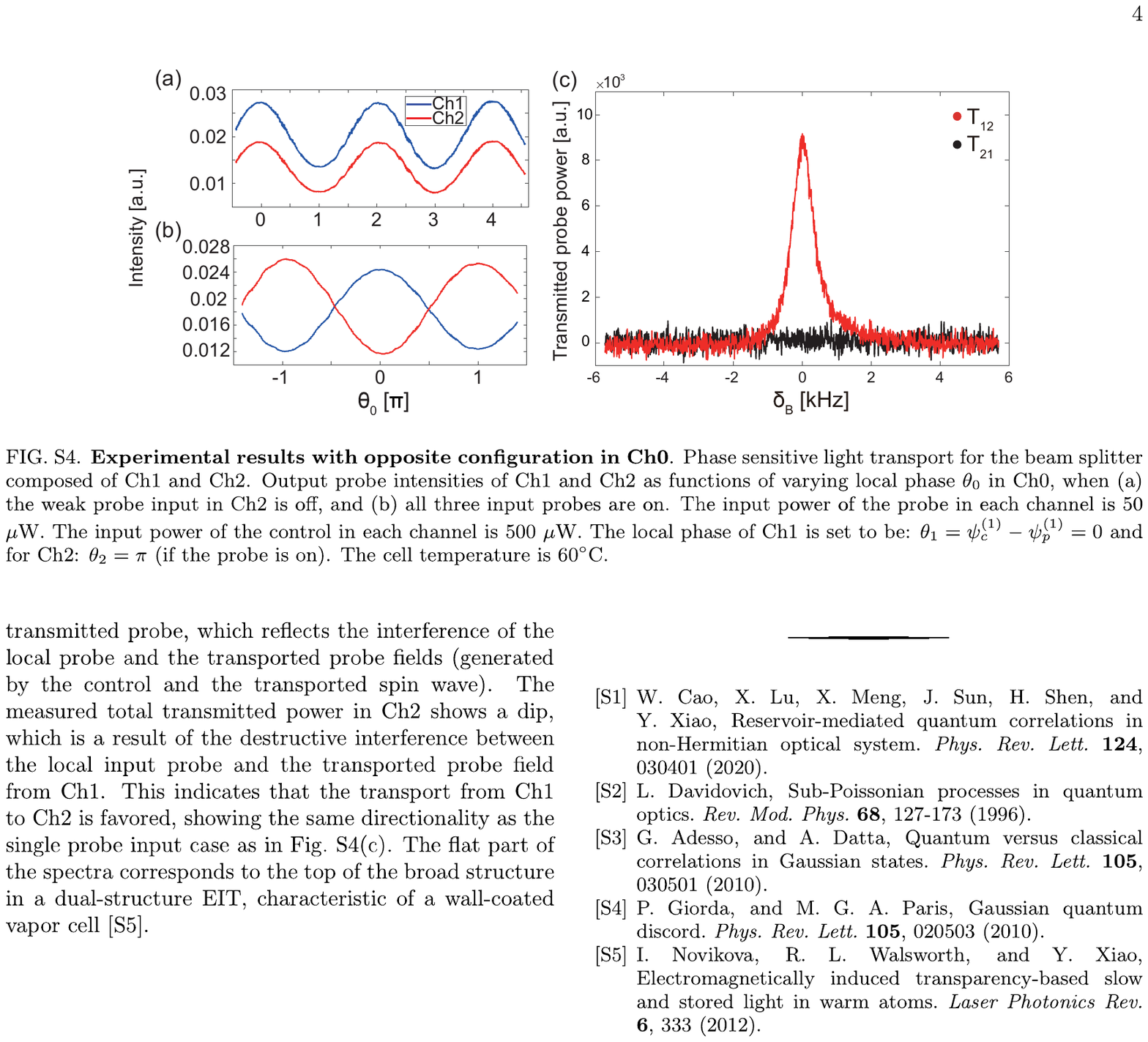}
\end{figure}

\begin{figure}
\centering
\includegraphics[width=1\textwidth]{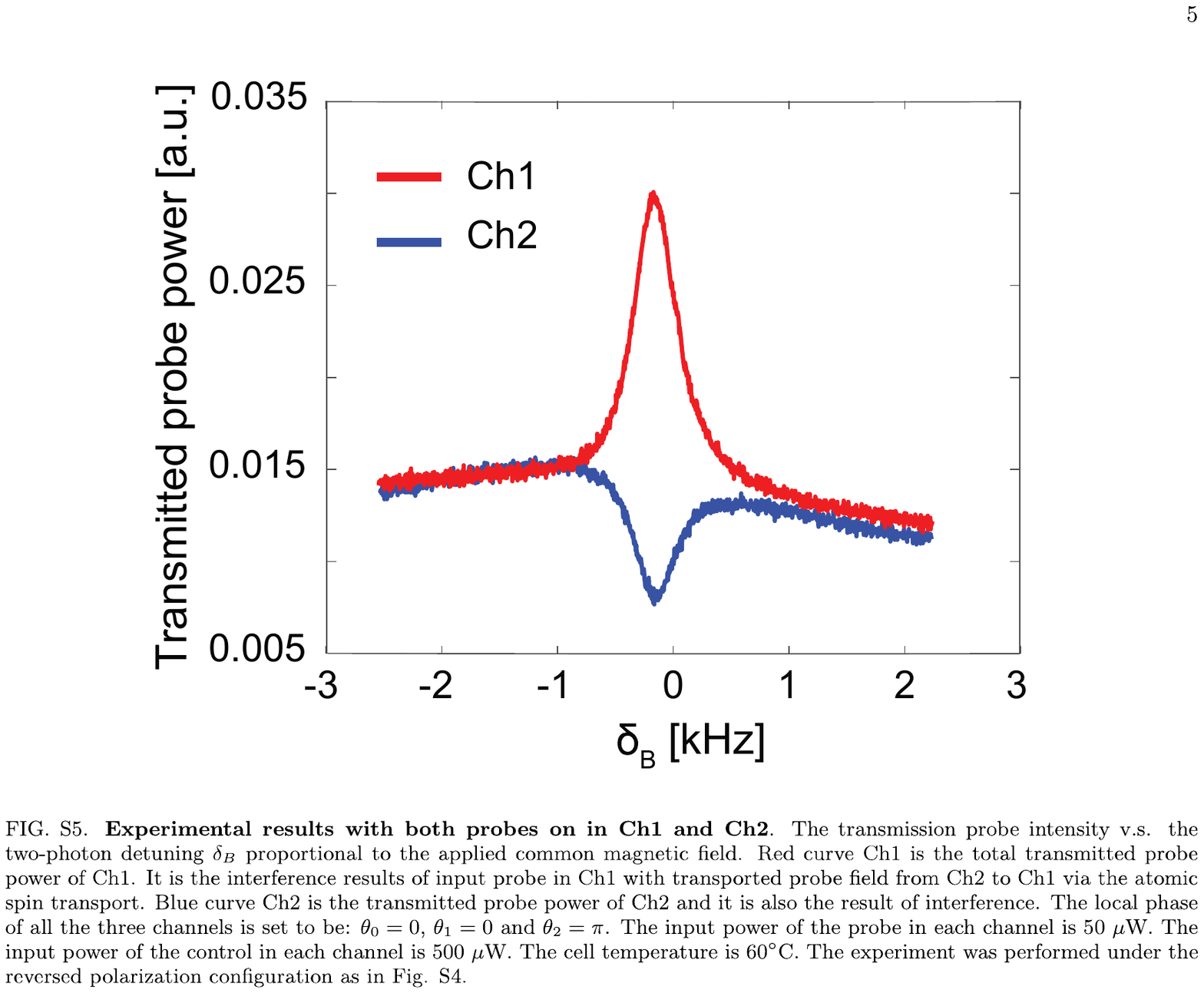}
\end{figure}
\end{document}